\documentclass[letterpaper]{article} % DO NOT CHANGE THIS
\usepackage[draft]{aaai2027}  % DO NOT CHANGE THIS
\usepackage[hyphens]{url}  % DO NOT CHANGE THIS
\usepackage{graphicx} % DO NOT CHANGE THIS
\usepackage{natbib}  % DO NOT CHANGE THIS AND DO NOT ADD ANY OPTIONS TO IT
\usepackage{caption} % DO NOT CHANGE THIS AND DO NOT ADD ANY OPTIONS TO IT
\usepackage{algorithm}
\usepackage{algorithmic}

\usepackage{newfloat}
\usepackage{listings}
\DeclareCaptionStyle{ruled}{labelfont=normalfont,labelsep=colon,strut=off} % DO NOT CHANGE THIS
\floatstyle{ruled}
\newfloat{listing}{tb}{lst}{}
\floatname{listing}{Listing}

\usepackage{booktabs}

\title{SongSculpt: Plan-Conditioned Fine-Grained Song Generation \\ via Structured Temporal Planning}
\author{
    Fan Wang\textsuperscript{\rm 1}\equalcontrib,
    Yutang Feng\textsuperscript{\rm 2}\equalcontrib,
    Shun Lei\textsuperscript{\rm 1},
    Jiatao Chen\textsuperscript{\rm 2},
    Mingyu Zhao\textsuperscript{\rm 1},
    Dapeng Wu\textsuperscript{\rm 1},
    Zijian Lin\textsuperscript{\rm 1},
    Jinchao Zhang\textsuperscript{\rm 2}\corresponding,
    Zhiyong Wu\textsuperscript{\rm 1}\corresponding
}

\affiliations{
    \textsuperscript{\rm 1}Shenzhen International Graduate School, Tsinghua University, Shenzhen, China\\
    \textsuperscript{\rm 2}WeChat AI, Tencent, China\\
    wang-f25@mails.tsinghua.edu.cn,
    yutangfeng@tencent.com,
    dayerzhang@tencent.com,
    zywu@sz.tsinghua.edu.cn
}

\begin{document}

\maketitle

\begin{abstract}

Recent advances in large language models have substantially improved long-form song generation. However, existing systems predominantly adopt a \emph{prompt-conditioned} paradigm, where heterogeneous musical control signals across different temporal granularities remain entangled within free-form prompts. Although recent methods improve controllability through temporal or section-level prompting, user intentions are still represented as unstructured textual descriptions, making fine-grained musical decisions difficult to explicitly organize, inspect, and revise before generation. To address this limitation, we propose \textbf{SongSculpt}, a \emph{plan-conditioned} framework for controllable long-form song generation. Instead of directly conditioning autoregressive generation on free-form prompts, SongSculpt first transforms user instructions into a \emph{Structured Temporal Plan}, which explicitly disentangles heterogeneous control signals into structured global attributes and temporally evolving segment-level plans before generation. The resulting plan is jointly grounded into autoregressive generation through segment-wise plan grounding, track-factorized realization, and boundary-aware transition modeling, while \textbf{SAMP-DPO} performs segment-aware multi-granularity preference optimization aligned with the hierarchical planning structure. Experiments demonstrate that SongSculpt consistently outperforms state-of-the-art open-source song generation models in both objective evaluations and human preference studies. More importantly, our results suggest that explicitly separating musical planning from musical realization establishes \emph{plan-conditioned generation} as a more effective conditioning paradigm for fine-grained controllable generation.

% 大语言模型的最新进展显著提升了长时程歌曲生成的水平。然而，现有系统大多采用 以提示词为条件（prompt-conditioned）的范式，跨越不同时间粒度的异构音乐控制信号依然纠缠于自由形式的提示词之中。尽管近期方法通过时序或段落级提示提升了可控性，用户意图仍被表征为非结构化的文本描述，使得细粒度的音乐决策难以在生成之前得到显式的组织、检查与修改。

% 为弥补这一局限，本文提出了 SongSculpt ——一个面向可控长时程歌曲生成的 以规划为条件（plan-conditioned）框架。SongSculpt 并不将自回归生成直接建立在自由形式提示词的条件之上，而是首先将用户指令转换为 结构化时序规划（Structured Temporal Plan），在生成之前将异构控制信号显式地解耦为结构化的全局属性与随时间演化的段落级规划。所得规划随后通过逐段落的规划落地、音轨分解式实现与边界感知的过渡建模，协同落地于自回归生成过程；与此同时，SAMP-DPO 执行与层次化规划结构相对齐的段落感知多粒度偏好优化。

% 实验表明，SongSculpt 在客观评测与人工主观实验中均持续优于最先进的开源歌曲生成模型。更为重要的是，本文的结果表明：将音乐规划与音乐实现显式分离，使 规划条件化生成（plan-conditioned generation）成为细粒度可控长时程歌曲生成的一种更为有效的条件化范式。

\end{abstract}


\section{Introduction}

Recent advances in large language models (LLMs)~\cite{brown2020language} and neural audio codecs~\cite{zeghidour2021soundstream,defossez2022high} have substantially advanced song generation~\cite{dhariwal2020jukebox,agostinelli2023musiclm,copet2024musicgen,yuan2025yue,ning2025diffrhythm,lei2025levo}, enabling unified autoregressive generation of lyrics, singing voices, accompaniment, and long-form musical structures directly from natural language descriptions. As generation quality continues to improve, the research focus is shifting from synthesizing plausible songs toward faithfully realizing users' creative intentions through fine-grained controllable song generation.

% 大语言模型（large language models, LLMs）与神经音频编解码器的最新进展极大地推动了歌曲生成技术的发展，使直接基于自然语言描述统一地自回归生成歌词、歌声、伴奏与长时程音乐结构成为可能。随着生成质量的持续提升，研究重心正从"合成听感合理的歌曲"转向"通过细粒度可控的歌曲生成，忠实地实现用户的创作意图"。

Existing systems have begun introducing temporal prompts and section-level conditions to improve controllability~\cite{lei2025levo,muse2025,wang2026segtune}. Nevertheless, they remain fundamentally prompt-conditioned, where heterogeneous musical control signals across multiple temporal granularities are expressed as free-form textual descriptions. Although such prompts provide flexible natural-language interfaces, they offer relatively weak structural organization, making individual planning decisions less interpretable, less directly editable, and difficult to revise independently. Consequently, autoregressive generation must still organize intertwined control decisions during realization, limiting structured fine-grained control over long-form musical development.

% 现有系统已开始引入时序提示与段落级条件以提升可控性。然而，它们在本质上仍是以提示词为条件（prompt-conditioned）的：跨越多重时间粒度的异构音乐控制信号均以自由形式的文本描述来表达。尽管此类提示提供了灵活的自然语言接口，其结构化组织能力却相对薄弱，使得单项规划决策的可解释性不足、难以直接编辑，且无法独立修改。因此，自回归生成仍须在实现过程中组织相互交织的控制决策，从而限制了对长时程音乐发展的结构化细粒度控制。

\begin{figure*}[t]
\centering
\includegraphics[width=0.9\textwidth]{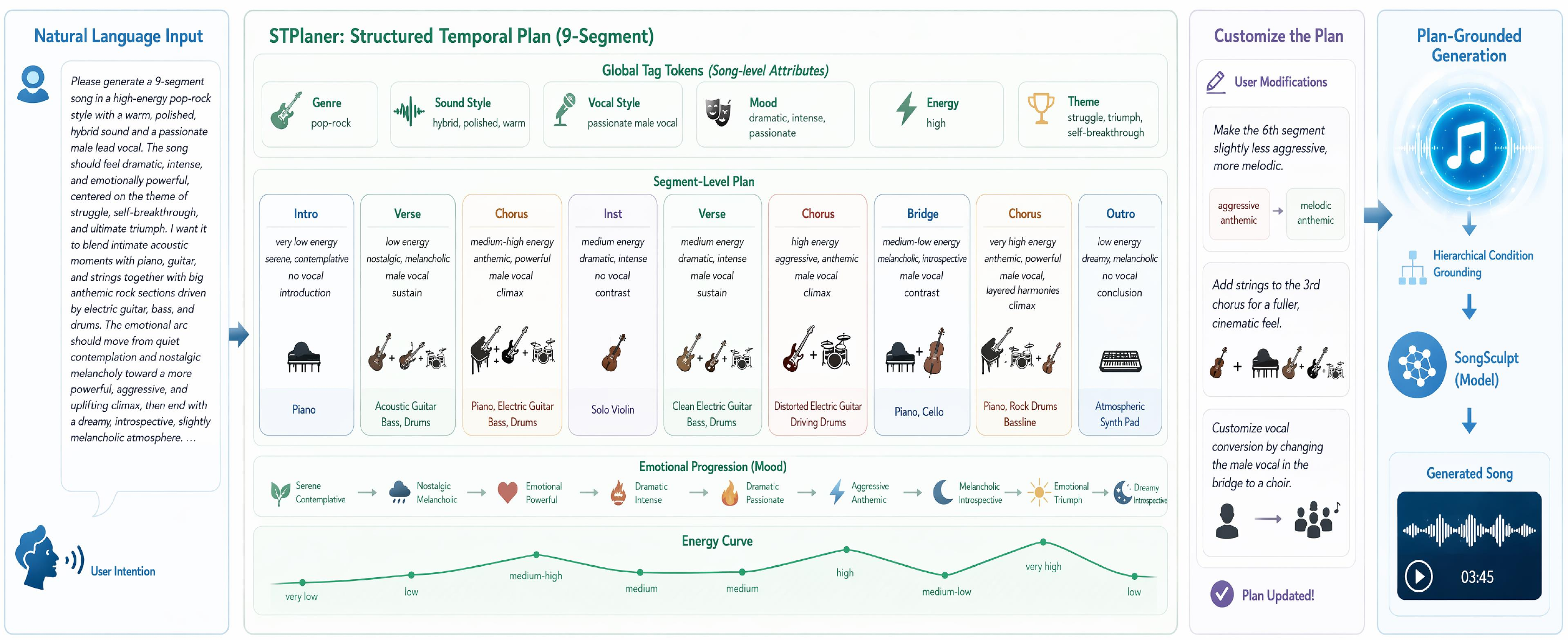}
\caption{From prompt-conditioned generation to plan-conditioned generation. Instead of directly conditioning autoregressive generation on free-form user instructions, SongSculpt first organizes user intentions into a Structured Temporal Plan that can be inspected and refined before guiding long-form song generation.}
\label{plan-conditioned_overview}
\end{figure*}

% 从"以提示词为条件"的生成到"以规划为条件"的生成。SongSculpt 并不将自回归生成直接建立在自由形式用户指令的条件之上，而是首先将用户意图组织为结构化时序规划——该规划在指导长时程歌曲生成之前，可经检查与细化。

Motivated by this observation, we revisit the conditioning paradigm of controllable song generation and propose \emph{plan-conditioned generation}. As illustrated in Figure~\ref{plan-conditioned_overview}, instead of directly conditioning autoregressive generation on free-form prompts, user intentions are first transformed into a \emph{Structured Temporal Plan}, where heterogeneous control signals are explicitly disentangled into structured global attributes and segment-level musical evolution. Unlike free-form prompts, the resulting plan serves as an interpretable and addressable intermediate representation that can be inspected and refined before generation, allowing users to revise individual planning decisions without repeatedly rewriting textual prompts.

% 基于上述观察，本文重新审视了可控歌曲生成的条件化范式，并提出 规划条件化生成（plan-conditioned generation）。如图 1 所示，本文并不将自回归生成直接建立在自由形式提示词的条件之上，而是首先将用户意图转换为 \emph{结构化时序规划}（Structured Temporal Plan），其中异构控制信号被显式地解耦为结构化的全局属性与片段级的音乐演进。与自由形式的提示词不同，所得规划充当一种可解释、可寻址的中间表征，可在生成之前接受检查与细化，使用户无需反复改写文本提示词，即可对单项规划决策进行修改。

Based on this paradigm, we propose \textbf{SongSculpt}, a plan-conditioned framework for fine-grained controllable song generation. SongSculpt explicitly separates \emph{musical planning} from \emph{musical realization}. Structured plans are jointly realized through three complementary components: Segment-wise Plan Grounding establishes explicit alignment between planning decisions and song segments, Track-Factorized Realization injects structured plans into track-aware generation, and Boundary-aware Transition Modeling enhances continuity across neighboring song sections. Together, these components enable structured, temporally aligned, and fine-grained realization of evolving musical intentions.

% 基于这一范式，本文提出了 SongSculpt ——一个面向细粒度可控歌曲生成的规划条件化框架。SongSculpt 将 音乐规划（musical planning）与 音乐实现（musical realization）显式分离。结构化规划通过三个功能互补的组件加以协同实现：逐段落的规划落地（Segment-wise Plan Grounding）在规划决策与歌曲段落之间建立显式对齐；音轨分解式实现（Track-Factorized Realization）将结构化规划注入音轨感知的生成过程；边界感知的过渡建模（Boundary-aware Transition Modeling）则增强相邻歌曲段落之间的连续性。这些组件相互配合，共同支撑了对不断演化的音乐意图的结构化、时序对齐且细粒度的实现。

Furthermore, we introduce \textbf{SAMP-DPO}, a segment-anchored multi-preference optimization framework designed to match the hierarchical organization of Structured Temporal Plan. Rather than optimizing entire songs uniformly, SAMP-DPO performs localized preference alignment over song-, boundary-, and segment-level regions, providing temporally localized supervision that complements the proposed plan-conditioned generation framework.

% 此外，本文提出了 SAMP-DPO——一种以段落为锚点的多偏好优化框架，其设计旨在匹配结构化时序规划的层次化组织方式。SAMP-DPO 并不对整首歌曲进行无差别的均匀优化，而是在歌曲级、边界级与片段级区域上执行局部化的偏好对齐，提供时间上局部化的监督，从而与所提出的规划条件化生成框架形成互补。

More broadly, SongSculpt demonstrates that explicitly separating musical planning from musical realization establishes plan-conditioned generation as an effective conditioning paradigm for fine-grained controllable song generation. Experiments show consistent improvements over state-of-the-art open-source song generation systems in both objective evaluations and human preference studies.

% 从更广泛的层面来看，SongSculpt 表明：将音乐规划与音乐实现显式分离，使规划条件化生成成为细粒度可控歌曲生成的一种有效条件化范式。实验表明，SongSculpt 在客观评测与人工偏好研究中均持续优于最先进的开源歌曲生成系统。

Our contributions are summarized as follows:

\begin{itemize}

\item We revisit the conditioning paradigm of controllable long-form song generation and formulate \emph{plan-conditioned generation}, which replaces entangled free-form prompts with an interpretable Structured Temporal Plan for structured, inspectable, and editable musical control.

% 本文重新审视了可控长时程歌曲生成的条件化范式，并提出了 规划条件化生成（plan-conditioned generation）——以可解释的结构化时序规划取代纠缠的自由形式提示词，实现结构化、可检查且可编辑的音乐控制。

\item We propose \textbf{SongSculpt}, a unified plan-conditioned framework that systematically transforms Structured Temporal Plans into controllable autoregressive song generation, enabling structured, temporally aligned, and fine-grained realization of user intentions.

% 本文提出了 SongSculpt——一个统一的规划条件化框架，系统地将结构化时序规划转化为可控的自回归歌曲生成，从而以结构化、时序对齐且细粒度的方式实现用户意图。

\item We propose \textbf{SAMP-DPO}, a segment-anchored multi-preference optimization framework that aligns preference supervision with the hierarchical structure of Structured Temporal Plans through localized optimization over song-, boundary-, and segment-level regions.

% 本文提出了 SAMP-DPO——一种以片段为锚点的多偏好优化框架，通过在歌曲级、边界级与片段级区域上执行局部化优化，使偏好监督与结构化时序规划的层次化结构相对齐。

\item Experiments demonstrate that SongSculpt consistently outperforms state-of-the-art open-source song generation models in both objective evaluations and human preference studies, validating the effectiveness of plan-conditioned generation for controllable long-form song generation.

% 实验表明，SongSculpt 在客观评测与人工偏好研究中均持续优于最先进的开源歌曲生成模型，验证了规划条件化生成对于可控长时程歌曲生成的有效性。

\end{itemize}

\begin{figure*}[t]
\centering
\includegraphics[width=0.9\textwidth]{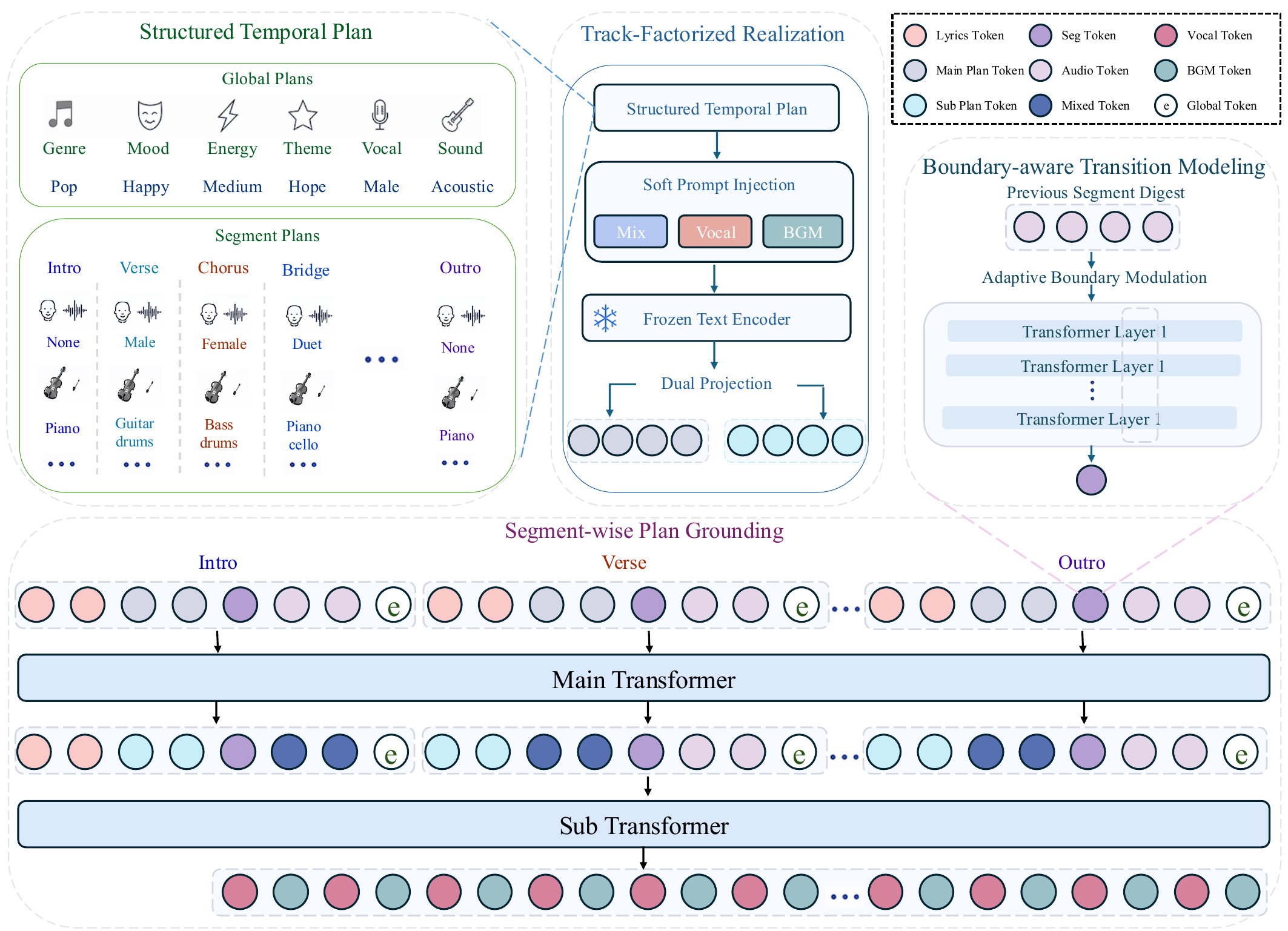}
\caption{
Overview of \textbf{SongSculpt}. Given a user instruction, SongSculpt first constructs a \emph{Structured Temporal Plan}, an interpretable intermediate representation that organizes musical intentions into structured global attributes and segment-level plans. The plan is subsequently realized through three complementary components: Segment-wise Plan Grounding, Track-Factorized Realization, and Boundary-aware Transition Modeling. Finally, SAMP-DPO performs plan-aligned preference optimization using segment-, boundary-, and song-level supervision.
}
\label{SongSculpt}
\end{figure*}

% SongSculpt 框架总览。给定用户指令，SongSculpt 首先构建 结构化时序规划（Structured Temporal Plan）——一种可解释的中间表征，将音乐意图组织为结构化的全局属性与片段级规划；随后，该规划通过三个功能互补的组件加以实现：逐片段的规划落地（Segment-wise Plan Grounding）、音轨分解式实现（Track-Factorized Realization）与边界感知的过渡建模（Boundary-aware Transition Modeling）；最后，SAMP-DPO 利用片段级、边界级与歌曲级监督，执行与规划相对齐的偏好优化。

\section{Related Work}

\subsection{Long-form Song Generation}

Recent LLM-based song generation systems such as Muse~\cite{muse2025} and LeVo~\cite{lei2025levo} enable autoregressive generation of complete songs with coherent vocals, accompaniment, and long-range musical structures. These methods primarily improve generation quality through stronger architectures, larger training corpora, and preference optimization~\cite{ouyang2022training,rafailov2023dpo,cideron2024musicrl}, while relying on free-form prompts or coarse attribute tags to specify user intentions before autoregressive generation, leaving the organization of temporally evolving musical decisions largely implicit. In contrast, SongSculpt reformulates controllable song generation as a plan-conditioned problem, explicitly organizing musical decisions before realization.

% 近年来，基于大语言模型（LLM）的歌曲生成系统（如 Muse 与 LeVo）已能够以自回归方式生成完整的歌曲，呈现出连贯的人声与伴奏以及长程的音乐结构。这类方法主要通过更强的模型架构、更大规模的训练语料以及偏好优化来提升生成质量；然而，它们在自回归生成之前仅依赖自由形式的提示或粗粒度的属性标签来指定用户意图，使得随时间演化的音乐决策的组织在很大程度上仍是隐式的。与之不同，SongSculpt 将可控歌曲生成重新表述为一个计划条件化问题（plan-conditioned problem），在最终音乐实现之前显式地组织各项音乐决策。

%%%%%%%%%%%%%%%%%%%%%%%%%%%%%%%%%%%%%%%%%%%%%%%%%%%%%%%

\subsection{Controllable Song Generation}

Controllable song generation has evolved from explicit musical attribute control~\cite{agostinelli2023musiclm,copet2024musicgen,wu2024musiccontrolnet,seedmusic2024} to temporally organized prompting. Recent systems such as Muse~\cite{muse2025} and SegTune~\cite{wang2026segtune} organize global and segment-level prompts to better describe musical evolution. However, heterogeneous control signals with different temporal scopes remain coupled within textual prompts, making them difficult to inspect and revise before generation, while also requiring the autoregressive model to implicitly organize and disentangle these mixed control signals during generation. Instead, SongSculpt formulates controllable song generation as a \emph{plan-conditioned} generation problem, where user intentions are first organized into an explicit Structured Temporal Plan that is structured, interpretable, editable, and naturally aligned with autoregressive generation.

% 可控歌曲生成已经从早期的显式音乐属性控制发展到时序化 Prompt 控制。近期代表性工作（如 Muse 和 SegTune）通过组织全局与段落级 Prompt 来描述歌曲的时间演进。然而，不同时间尺度的异构控制信息仍然耦合于文本 Prompt 中，不仅难以在生成前进行检查、修改与逐步优化，而且需要自回归模型在生成过程中自行理解、组织并解耦这些混合控制信号。与之不同，SongSculpt 将可控歌曲生成重新定义为一种 plan-conditioned generation：用户意图首先被组织为显式的 Structured Temporal Plan，在自回归生成之前即具备结构化、可解释、可编辑，并天然适配生成过程的规划形式。

%%%%%%%%%%%%%%%%%%%%%%%%%%%%%%%%%%%%%%%%%%%%%%%%%%%%%%%

\subsection{Preference Alignment for Song Generation}

Preference optimization has become an effective paradigm~\cite{ouyang2022training,rafailov2023dpo} for improving song generation quality and instruction following. Existing methods primarily optimize whole-song preferences~\cite{cideron2024musicrl,lei2025levo}, providing a single supervisory signal for the entire generation process. While effective for improving overall quality, such supervision is not naturally aligned with temporally localized control, making it difficult to selectively optimize individual song sections or transitions. To complement plan-conditioned generation, we propose Segment-Anchored Multi-Preference Optimization (SAMP-DPO), which aligns preference supervision with the hierarchical structure of Structured Temporal Plans through segment-, boundary-, and song-level optimization. By matching supervision granularity with planning granularity, SAMP-DPO provides more localized and structurally consistent preference signals for controllable long-form song generation.

% 偏好优化已成为提升歌曲生成质量和指令遵循能力的重要范式。现有方法主要基于整曲偏好进行优化，为整个生成过程提供统一监督。尽管能够有效提升整体质量，但这种监督方式与时间局部化控制并不天然匹配，因此难以针对单个歌曲段落或段间过渡进行选择性优化。为与 plan-conditioned generation 相配套，本文提出 Segment-Anchored Multi-Preference Optimization（SAMP-DPO），通过段落级、边界级和整曲级三个粒度，使偏好监督与 Structured Temporal Plan 的层次结构保持一致。通过使监督粒度与规划粒度相匹配，SAMP-DPO 为可控长歌曲生成提供更加局部化且结构一致的偏好信号。

%%%%%%%%%%%%%%%%%%%%%%%%%%%%%%%%%%%%%%%%%%%%%%%%%%%%%%%%%%%%%%%%%%%%%%%%%%%%%%%
\section{Method}

\subsection{Overview}

Figure~\ref{SongSculpt} illustrates the overall framework of \textbf{SongSculpt}, built upon a re-implemented and independently trained LeVo~\cite{lei2025levo} autoregressive backbone. Given a natural-language instruction, SongSculpt first constructs a \emph{Structured Temporal Plan}, which organizes musical intentions into structured global attributes and temporally ordered segment-level plans, serving as the unified control representation throughout generation.

% 图 2 展示了 SongSculpt 的整体框架，该框架构建于复现并独立训练的 LeVo 自回归骨干网络之上。给定自然语言指令，SongSculpt 首先构建 结构化时序规划（Structured Temporal Plan），将音乐意图组织为结构化的全局属性与时序排列的片段级规划，作为贯穿整个生成过程的统一控制表征。

The Structured Temporal Plan is jointly realized by three complementary components. Segment-wise Plan Grounding establishes explicit alignment between planning decisions and autoregressive generation. Track-Factorized Realization translates structured plans into track-aware conditioning for mixed-track, vocal-track and accomplishment-track generation. Boundary-aware Transition Modeling improves continuity across neighboring song segments through explicit transition modeling. Finally, SAMP-DPO performs localized preference optimization that naturally follows the hierarchical organization of the Structured Temporal Plan.

% 结构化时序规划由三个功能互补的组件协同实现：逐片段的规划落地（Segment-wise Plan Grounding）在规划决策与自回归生成之间建立显式对齐；音轨分解式实现（Track-Factorized Realization）将结构化规划转化为面向混合音轨、人声轨和伴奏轨生成的音轨感知条件；边界感知的过渡建模（Boundary-aware Transition Modeling）通过显式的过渡建模改善相邻歌曲片段之间的连续性。最后，SAMP-DPO 执行局部化的偏好优化，其优化方式自然遵循结构化时序规划的层次化组织。
%%%%%%%%%%%%%%%%%%%%%%%%%%%%%%%%%%%%%%%%%%%%%%%%%%%%%%%%%%%%%%%%%%%%%%%%%%%%%%%

\subsection{Structured Temporal Planning}

Instead of directly conditioning autoregressive generation on free-form textual descriptions, SongSculpt first transforms user instructions into an interpretable intermediate representation termed the \emph{Structured Temporal Plan}. This planning stage explicitly organizes heterogeneous musical intentions into structured control variables before generation, separating musical planning from musical realization while providing a unified representation for subsequent generation and optimization.

% SongSculpt 并不将自回归生成直接建立在自由形式文本描述的条件之上，而是首先将用户指令转化为一种可解释的中间表征，称为 结构化时序规划（Structured Temporal Plan）。这一规划阶段在生成之前将异构的音乐意图显式地组织为结构化的控制变量，既将音乐规划与音乐实现相分离，又为后续的生成与优化提供了统一的表征。

Formally, given a user instruction $\mathcal{X}$, the planning model predicts
\begin{equation}
\mathcal{P}
=
(\mathcal{G},\mathcal{S}),
\end{equation}
where $\mathcal{G}$ denotes globally shared song attributes, and
\begin{equation}
\mathcal{S}
=
\{
\mathbf{s}_1,
\mathbf{s}_2,
\ldots,
\mathbf{s}_N
\}
\end{equation}
contains the ordered plans of all $N$ song segments.

The global plan is defined as
\begin{equation}
\mathcal{G}
=
\{
g_{\mathrm{genre}},
g_{\mathrm{sound}},
g_{\mathrm{mood}},
g_{\mathrm{energy}},
g_{\mathrm{theme}},
g_{\mathrm{vocal}}
\},
\end{equation}
which specifies globally consistent musical characteristics shared across the entire song.

Each segment plan is represented as
\begin{equation}
\mathbf{s}_i
=
\{
e,
ec,
re,
m,
f,
ss,
v,
d,
inst
\},
\end{equation}
where the attributes describe segment energy, energy change, relative energy, mood, structural function, sound style, vocal presence, arrangement density, and instrumentation, respectively. Unlike conventional prompts that express temporal evolution implicitly in free-form language, the Structured Temporal Plan represents each musical decision as an explicit and independently addressable control variable while preserving its temporal organization throughout the song. The complete serialization format is provided in Appendix~A.

% 其中，各属性分别描述片段能量、能量变化、相对能量、情绪、结构功能、声音风格、人声存在与否、编曲密度与配器。与以自由形式语言隐式表达时序演化的传统提示词不同，结构化时序规划将每一项音乐决策表征为显式且可独立寻址的控制变量，同时保留其在整首歌曲中的时序组织。完整的序列化格式见附录~A。

To train the planner, we fine-tune Qwen3.5-4B~\cite{yang2025qwen3} through a two-stage curriculum~\cite{bengio2009curriculum}. The first stage uses temporally ordered segment captions to learn explicit structural evolution across song sections. The second stage further aligns the planner with original whole-song captions, enabling direct prediction of structured plans from unconstrained user instructions.

% 为训练规划器，本文通过两阶段的课程（two-stage curriculum）对 Qwen3.5-4B 进行微调。第一阶段利用按时间排序的片段描述，学习跨歌曲段落的显式结构演化；第二阶段进一步使规划器与原始的整曲描述相对齐，从而使规划器能够根据不受约束的用户指令直接预测结构化规划。

During both stages, each training caption is prepended with
\begin{center}
\emph{"Please generate a $N$-segment song ..."},
\end{center}
where $N$ specifies the desired number of song segments. Rather than annotating a predefined structure, the planner jointly infers musical content and song organization conditioned on the target segment count, allowing song structure itself to become a controllable planning variable.

% 其中，N 指定期望的歌曲片段数量。规划器并非对预定义的结构进行标注，而是以目标片段数为条件，联合推断音乐内容与歌曲组织方式，从而使歌曲结构本身成为一种可控的规划变量。

The resulting Structured Temporal Plan serves as the unified control representation for the subsequent generation process, allowing users to refine planning decisions before autoregressive realization.

% 所得的结构化时序规划将作为后续生成过程的统一控制表征，使用户能够在自回归实现之前对规划决策进行细化。

%%%%%%%%%%%%%%%%%%%%%%%%%%%%%%%%%%%%%%%%%%%%%%%%%%%%%%%%%%%%%%%%%%%%%%%%%%%%%%%

\begin{table*}[t]
\centering
\small
\setlength{\tabcolsep}{4.5pt}
\renewcommand{\arraystretch}{1.15}
\setlength{\aboverulesep}{0pt}
\setlength{\belowrulesep}{0pt} 

\begin{tabular}{lcccccc|cccccccc}
\toprule

&
\multicolumn{5}{c}{\textbf{AudioBox}}
&
&
\multicolumn{8}{c}{\textbf{SongBench}}
\\

\cmidrule(lr){2-6}
\cmidrule(l){8-15}

Method
& CE & CU & PC & PQ & Mean & PER$\downarrow$
& Melody & Arrange. & Musicality & Vocal & Instr. & Mixing & Struct. & Mean
\\

\midrule

\multicolumn{15}{c}{\textit{Commercial Model}}\\
\midrule

Suno V5.5
&
7.52 & 7.91 & 6.11 & 6.18 & 7.43 & 63.6
&
\textbf{6.82} & 7.21 & \underline{5.91}
& \textbf{6.95} & 7.14 & \textbf{6.97}
& \textbf{7.03} & \underline{6.86}
\\

\midrule

\multicolumn{15}{c}{\textit{Open-source Models}}\\
\midrule

DiffRhythm 2
&
6.59 & 7.06 & 5.95 & 7.27 & 6.72 & 14.2
&
4.98 & 4.76 & 3.98 & 5.08 & 4.61 & 4.43 & 4.95 & 4.68
\\

YuE
&
7.10 & 7.66 & 5.11 & 7.87 & 6.93 & 28.9
&
4.98 & 5.05 & 3.90 & 4.95 & 5.50 & 4.87 & 4.75 & 4.86
\\

Muse
&
7.50 & 7.72 & 6.39 & 8.10 & 7.42 & 147.4
&
5.96 & 6.04 & 5.00 & 6.35 & 6.20 & 5.97 & 5.84 & 5.91
\\

HeartMuLa
&
7.63 & 7.90 & 6.11 & 8.23 & 7.47 & 33.7
&
6.27 & 6.34 & 5.33 & 6.62 & 6.53 & 6.33 & 6.20 & 6.23
\\

AceStep 1.5
&
7.57 & 7.88 & 6.28 & 8.23 & 7.49 & 19.8
&
6.42 & 6.70 & 5.33 & 6.19 & 6.22 & 6.41 & 6.34 & 6.29
\\

LeVo 2
&
7.69 & 8.03 & \underline{6.45} & 8.37
& 7.63 & 22.2
&
6.70 & \underline{7.22} & 5.81 & \underline{6.89}
& \underline{7.25} & 6.91 & 6.58 & 6.77
\\

\midrule

SongSculpt
&
\textbf{7.84} & \underline{8.05} & \textbf{6.49} & \underline{8.41}
& \textbf{7.70} & \textbf{4.5}
&
\underline{6.80} & \textbf{7.40} & \textbf{6.05}
& \underline{6.89} & \textbf{7.38} & \underline{6.96}
& \underline{6.66} & \textbf{6.88}
\\

\quad w/o DPO
&
\underline{7.79} & \textbf{8.06} & 6.33 & \textbf{8.49}
& \underline{7.66} & \underline{8.6}
&
6.20 & 6.75 & 5.28 & 6.28 & 6.89 & 6.36 & 6.04 & 6.26
\\

\bottomrule
\end{tabular}

\caption{
Comparison of objective evaluation results.
AudioBox evaluates perceptual audio quality, SongBench evaluates professional musical quality, and PER measures lyric pronunciation accuracy (lower is better).
``w/o DPO'' denotes the SongSculpt variant trained with SFT only.
Best and second-best results are shown in \textbf{bold} and \underline{underline}, respectively.
}

\label{tab:objective_results}

\end{table*}

\subsection{Plan-conditioned Song Generation}

Given the Structured Temporal Plan, SongSculpt realizes high-level musical planning through three complementary components: Segment-wise Plan Grounding establishes temporal alignment between planning and generation, Track-Factorized Realization translates structured plans into track-aware generation conditions, and Boundary-aware Transition Modeling enhances continuity across neighboring song segments.

% 给定结构化时序规划之后，SongSculpt 通过三个功能互补的组件实现高层的音乐规划：逐片段的规划落地（Segment-wise Plan Grounding）在规划与生成之间建立时序对齐；音轨分解式实现（Track-Factorized Realization）将结构化规划转化为音轨感知的生成条件；边界感知的过渡建模（Boundary-aware Transition Modeling）则增强相邻歌曲片段之间的连续性。

%%%%%%%%%%%%%%%%%%%%%%%%%%%%%%%%%%%%%%%%%%%%%%%%%%%%%%%%%%%%%%%%%%%%%%%%%%%%%%%

\smallskip
\noindent\textbf{Segment-wise Plan Grounding}

The Structured Temporal Plan specifies musical decisions at different temporal granularities. To preserve this organization during autoregressive generation, SongSculpt explicitly grounds planning information into an interleaved autoregressive sequence that maintains alignment between planning decisions and their corresponding song segments.

% 结构化时序规划在不同时间粒度上规定音乐决策。为在自回归生成过程中保持这一组织方式，SongSculpt 将规划信息显式地落地到交错式自回归序列之中，从而使规划决策与其对应的歌曲片段之间始终保持对齐。

Global tags, segment plans, and lyrics are independently tokenized before being interleaved with codec audio tokens, preserving semantic boundaries among different control sources while maintaining explicit alignment with musical structure.

% 全局标签、段落规划与歌词首先被独立地进行词元化，随后再与编解码器音频词元（codec audio tokens）交错排列——这一设计在保持与音乐结构显式对齐的同时，保留了不同控制来源之间的语义边界。

For each song segment, the grounded sequence is organized as
\begin{equation}
\mathbf{x}_i
=
[
\mathbf{t}_i,\;
\mathbf{l}_i,\;
\texttt{[SEG]},\;
\mathbf{a}_i
],
\end{equation}
where $\mathbf{t}_i$, $\mathbf{l}_i$, and $\mathbf{a}_i$ denote segment-tag tokens, lyric tokens, and codec audio tokens, respectively.

The complete autoregressive sequence becomes
\begin{equation}
\mathbf{X}
=
[
\mathbf{g},
\mathbf{x}_1,
\mathbf{x}_2,
\ldots,
\mathbf{x}_N
],
\end{equation}
where $\mathbf{g}$ denotes the global planning tokens shared across all song segments.

This explicit serialization establishes temporal alignment between structured planning information and the autoregressive generation process, allowing segment-level musical decisions to be grounded at their corresponding structural positions instead of being implicitly inferred throughout generation.

% 这种显式的序列化方式在结构化规划信息与自回归生成之间建立了时序对齐，使段落级的音乐决策得以在其对应的结构位置上直接落地，而无需在整个生成过程中被隐式推断。

%%%%%%%%%%%%%%%%%%%%%%%%%%%%%%%%%%%%%%%%%%%%%%%%%%%%%%%%%%%%%%%%%%%%%%%%%%%%%%%

\smallskip
\noindent\textbf{Track-Factorized Realization}

While Segment-wise Plan Grounding establishes where planning information is aligned with autoregressive generation, Track-Factorized Realization determines how structured plans condition different generation streams through a two-stage track-factorized conditioning strategy.

% 逐段落的规划落地（Segment-wise Plan Grounding）确立了规划信息与自回归生成相对齐的位置，而音轨分解式实现（Track-Factorized Realization）则通过两阶段的音轨分解式条件化策略，决定结构化规划如何对不同生成流进行条件化。

The grounded plan is first instantiated into three track-aware soft prompt templates (MIX, VOCAL, and BGM), which are encoded by the frozen Qwen3-Embedding-0.6B~\cite{zhang2025qwen3embedding} text encoder to produce a shared semantic representation $\mathbf{h}$.
% 落地后的规划首先被实例化为三个音轨感知的软提示模板（MIX、VOCAL 与 BGM），并由冻结的 Qwen3-Embedding-0.6B 文本编码器进行编码，从而产生共享语义表征 h。
\begin{equation}
\mathbf{h}_{\mathrm{main}}
=
W_{\mathrm{main}}\mathbf{h},
\qquad
\mathbf{h}_{\mathrm{sub}}
=
W_{\mathrm{sub}}\mathbf{h},
\end{equation}
where $W_{\mathrm{main}}$ and $W_{\mathrm{sub}}$ are learnable projection heads.

The two branches are injected into the two transformer streams inherited from the LeVo backbone, where the main branch conditions mixed-track generation (T1) and the sub branch conditions subsequent dual-track generation (T2). In this way, SongSculpt factorizes structured planning information at the conditioning level without modifying the underlying autoregressive architecture.

% 两个分支分别注入继承自 LeVo 骨干网络的两路 transformer 流之中：主分支对混合音轨生成（T1）进行条件化，子分支则对后续的双音轨生成（T2）进行条件化。通过这种方式，SongSculpt 在条件化层面实现了结构化规划信息的分解，而无需对底层的自回归架构作改动。

Within T2, a second factorization stage further decouples vocal and accompaniment generation. Both tracks share the same sub-branch representation while using independent codec embeddings and prediction heads,

% 在 T2 阶段内部，第二个分解阶段进一步解耦了人声与伴奏的生成：两条音轨共享同一子分支表征，但各自采用独立的编解码器嵌入（codec embeddings）与预测头（prediction heads），
\begin{equation}
\mathcal{L}_{\mathrm{dual}}
=
\mathcal{L}_{\mathrm{vocal}}
+
\mathcal{L}_{\mathrm{bgm}}.
\end{equation}
Overall, the first stage separates shared planning semantics into mixed-track and dual-track condition branches, while the second stage further specializes vocal and accompaniment realization through shared representations with track-specific prediction heads. This hierarchical factorization enables track-aware control while preserving a unified semantic representation throughout generation.

% 总体而言，第一阶段将共享的规划语义分解至混合音轨与双音轨两个条件分支，第二阶段则通过"共享表征＋音轨特定预测头"的设计进一步细化人声与伴奏的实现。这种层次化的分解机制在实现音轨感知控制的同时，于整个生成过程中始终保持了统一的语义表征。

%%%%%%%%%%%%%%%%%%%%%%%%%%%%%%%%%%%%%%%%%%%%%%%%%%%%%%%%%%%%%%%%%%%%%%%%%%%%%%%

\smallskip
\noindent\textbf{Boundary-aware Transition Modeling}

While Segment-wise Plan Grounding explicitly aligns planning decisions with individual song segments, it does not explicitly model how neighboring segments should transition during autoregressive generation. To complement structured planning with cross-segment continuity, SongSculpt further introduces a Boundary-aware Transition Modeling module following recent boundary-aware sequence modeling approach~\cite{boundary2025} while adapting it to plan-conditioned song generation.. 

% 尽管逐片段的规划落地在规划决策与各歌曲片段之间建立了显式对齐，却并未显式建模相邻片段在自回归生成过程中应如何过渡。为以跨片段的连续性补足结构化规划，SongSculpt 借鉴近期的边界感知序列建模方法，进一步引入了边界感知的过渡建模模块，并将其适配于规划条件化的歌曲生成。

For segment $i$, the boundary representation is summarized from the final $K$ codec tokens of the preceding mixed-track segment,
% 对于片段 i，其边界表征从前一混合音轨片段的最后 K 个编解码器词元中归纳得到，
\begin{equation}
\mathbf{z}_{i-1}
=
\frac{1}{K}
\sum_{k=1}^{K}
E(c_{i-1}^{(k)}),
\end{equation}
where $E(\cdot)$ denotes the codec embedding of the first codebook.

The shared boundary summary is reused across all transformer layers. At layer $l$, an independent modulation network~\cite{perez2018film} predicts
% 共享的边界摘要在所有 transformer 层中被复用。在第 l 层，一个独立的调制网络负责预测
\begin{equation}
\gamma^{(l)},
\;
\beta^{(l)}
=
f^{(l)}(\mathbf{z}_{i-1}),
\end{equation}
which modulates a learnable boundary prototype
$\mathbf{b}^{(l)}$,
\begin{equation}
\hat{\mathbf{s}}^{(l)}
=
(1+\gamma^{(l)})
\odot
\mathbf{b}^{(l)}
+
\beta^{(l)}.
\end{equation}
The resulting boundary representation is injected into the segment boundary during autoregressive generation, providing transition-aware contextual information while leaving the remaining generation process unchanged. In this way, Segment-wise Plan Grounding determines the realization of each individual segment, whereas Boundary-aware Transition Modeling explicitly strengthens continuity across neighboring song sections.

% 所得的边界表征在自回归生成过程中被注入片段边界处，提供过渡感知的上下文信息，同时不改变其余的生成过程。由此，逐片段的规划落地决定各单个片段的实现方式，而边界感知的过渡建模则显式地增强相邻歌曲段落之间的连续性。

%%%%%%%%%%%%%%%%%%%%%%%%%%%%%%%%%%%%%%%%%%%%%%%%%%%%%%%%%%%%%%%%%%%%%%%%%%%%%%%

\begin{table*}[t]
\centering
\small
\setlength{\tabcolsep}{6pt}
\renewcommand{\arraystretch}{1.15}

\begin{tabular}{lccccccc|c}
\toprule

&
\multicolumn{7}{c}{\textbf{Segment-level MuLan}}
&
\textbf{Boundary Continuity}
\\

\cmidrule(lr){2-8}
\cmidrule(l){9-9}

Method
& Intro
& Verse
& Chorus
& Inst.
& Bridge
& Outro
& Mean
& Continuity
\\

\midrule

Muse
& \textbf{0.21}
& 0.29
& 0.29
& \textbf{0.19}
& 0.22
& \textbf{0.19}
& 0.25
& 7.10
\\

SongSculpt
& \textbf{0.21}
& \textbf{0.38}
& \textbf{0.36}
& 0.17
& \textbf{0.31}
& \textbf{0.19}
& \textbf{0.30}
& \textbf{7.42}
\\

\bottomrule
\end{tabular}

\caption{
Fine-grained controllability comparison between Muse and SongSculpt. Segment-level MuLan similarity measures semantic consistency between each generated section and its corresponding target description. Continuity is measured by the MuLan cosine similarity within a 3-second window centered at adjacent segment boundaries. Higher values indicate better fine-grained controllability and smoother cross-segment transitions.
}

\label{tab:segment_results}

\end{table*}

\begin{table}[t]
\centering
\small
\setlength{\tabcolsep}{4.5pt}
\renewcommand{\arraystretch}{1.15}
\setlength{\aboverulesep}{0pt}
\setlength{\belowrulesep}{0pt} 

\begin{tabular}{lccccccc}
\toprule

Method
& Ovl. & V\&A & Har. & Struct. & Lyr. & IF. & Mean
\\

\midrule

\multicolumn{8}{c}{\textit{Commercial Model}}\\
\midrule

Suno V5.5
&
\textbf{4.14}
& \textbf{4.02}
& \textbf{4.16}
& \textbf{4.22}
& \textbf{4.33}
& \textbf{4.07}
& \textbf{4.16}
\\

\midrule

\multicolumn{8}{c}{\textit{Open-source Models}}\\
\midrule

DiffRhythm-2
&
2.77
& 2.81
& 2.88
& 3.04
& 3.55
& 2.73
& 2.96
\\

YuE
&
2.77
& 2.72
& 3.03
& 3.05
& 3.47
& 2.81
& 2.98
\\

Muse
&
3.10
& 3.17
& 3.38
& 3.29
& 3.22
& 3.01
& 3.19
\\

HeartMuLa
&
3.59
& 3.57
& 3.65
& 3.61
& 3.99
& 3.21
& 3.60
\\

AceStep 1.5
&
3.14
& 3.12
& 3.31
& 3.36
& 3.78
& 2.64
& 3.23
\\

LeVo 2
&
3.61
& 3.69
& 3.77
& 3.83
& 3.91
& 3.04
& 3.64
\\

\midrule

SongSculpt
&
\underline{3.88}
& \underline{3.88}
& \underline{3.98}
& \underline{4.01}
& \underline{4.17}
& \underline{3.76}
& \underline{3.95}
\\

\bottomrule

\end{tabular}

\caption{
Subjective evaluation results.
Ovl. denotes Overall Musical Quality,
V\&A denotes Vocal \& Accompaniment Quality,
Har. denotes Vocal–Accompaniment Harmony,
Struct. denotes Song Structure Clarity,
Lyr. denotes Lyric Accuracy,
and IF. Instruction Faithfulness.
Best and second-best results are shown in \textbf{bold} and \underline{underline}, respectively.
}

\label{tab:subjective_results}
\end{table}

\subsection{Segment-Anchored Multi-Preference Optimization}

The Structured Temporal Plan organizes supervision naturally into multiple temporal granularities. To match this hierarchical organization during optimization, we propose \textbf{Segment-Anchored Multi-Preference Optimization (SAMP-DPO)}, which performs localized preference optimization over three complementary supervision regions: individual song segments, neighboring segment boundaries, and complete songs.

% 结构化时序规划将监督信号自然地组织到多个时间粒度之中。为在优化过程中匹配这一层次化组织，本文提出了\textbf{以片段为锚点的多偏好优化}（Segment-Anchored Multi-Preference Optimization, SAMP-DPO），在三种功能互补的监督区域上执行局部化的偏好优化：单个歌曲片段、相邻片段边界与完整歌曲。

Rather than uniformly optimizing all autoregressive tokens, SAMP-DPO first samples a target supervision region
$R$
(segment, boundary, or whole song) and masks all remaining tokens. The log-probability over the selected region is computed using length-normalized averaging,
% SAMP-DPO 并不对所有自回归词元进行无差别的均匀优化，而是首先采样一个目标监督区域 R（片段、边界或整首歌曲），并掩蔽其余全部词元。所选区域上的对数概率采用长度归一化的平均方式计算，
\begin{equation}
\bar{\ell}_{\theta}(y|R)
=
\frac{1}{|R|}
\sum_{t\in R}
\log
\pi_{\theta}
(
y_t
|
x,
y_{<t}
),
\end{equation}
which prevents longer regions from dominating optimization and maintains comparable preference scales across different supervision granularities.

% 该设计可防止较长区域主导优化过程，并在不同监督粒度之间维持可比的偏好尺度。

The masked DPO objective is then formulated as
\begin{equation}
\mathcal{L}_{\mathrm{mask}}
=
-
\log
\sigma
\!\left(
\beta
\left[
\Delta_w
-
\Delta_l
\right]
\right),
\end{equation}
where
\begin{equation}
\Delta_w
=
\bar{\ell}_{\theta}(y_w|R)
-
\bar{\ell}_{\mathrm{ref}}(y_w|R),
\end{equation}
\begin{equation}
\Delta_l
=
\bar{\ell}_{\theta}(y_l|R)
-
\bar{\ell}_{\mathrm{ref}}(y_l|R),
\end{equation}
and $y_w$ and $y_l$ denote the preferred and rejected responses, respectively.

Training is performed on a mixed preference dataset containing segment-, boundary-, and song-level preference pairs. Instead of jointly optimizing all supervision granularities within every iteration, each optimization step samples a single supervision granularity, while the corresponding masked DPO loss is further weighted according to annotation confidence and preference quality.

% 训练在包含片段级、边界级与歌曲级偏好对的混合偏好数据集上进行。本文并不在每次迭代中联合优化所有监督粒度，而是在每个优化步骤中采样单一的监督粒度，并依据标注置信度与偏好质量对相应的掩码 DPO 损失进一步加权。

The three supervision granularities naturally complement one another. Segment-level supervision improves the realization fidelity of individual planning decisions by correcting localized generation errors. Boundary-level supervision directly optimizes transitions between neighboring song sections, complementing the proposed Boundary-aware Transition Modeling. Song-level supervision preserves overall musical quality throughout the complete composition. Consequently, the optimization process follows the same hierarchical organization as the Structured Temporal Plan, establishing a unified framework that consistently aligns planning, generation, and preference optimization.

% 三种监督粒度天然互补：片段级监督通过纠正局部化的生成错误，提升单项规划决策的实现保真度；边界级监督直接优化相邻歌曲段落之间的过渡，与所提出的边界感知过渡建模形成互补；歌曲级监督则保障完整作品的整体音乐质量。由此，优化过程遵循与结构化时序规划相同的层次化组织，建立起一个将规划、生成与偏好优化一致对齐的统一框架。

\section{Experiments}

\subsection{Datasets}

SongSculpt is trained on an internal corpus of approximately 1.3M songs. Each song is first segmented by SongFormer~\cite{songformer2025}, followed by lyric transcription and caption generation using AceStep-1.5~\cite{gong2025acestep}. Based on the resulting captions, Qwen3.5-4B predicts disentangled global planning attributes from the whole-song description and structured segment-level attributes from each segment conditioned on the corresponding global plan.

% SongSculpt 基于约 130 万首歌曲组成的内部数据集进行训练。每首歌曲首先由 SongFormer 完成结构切分，随后利用 AceStep-1.5 完成歌词转录与描述文本生成。在此基础上，Qwen3.5-4B 从整曲描述中预测解耦的全局规划属性，并结合对应的全局规划信息，从各段描述中推断结构化的段落级属性，最终构建用于训练与生成的 Structured Temporal Plans。

To train SAMP-DPO, we further construct a localized multi-granularity preference dataset. For each song, eight stochastic candidates are evaluated using SongBench~\cite{songbench2025}, segment-level MuLan similarity~\cite{huang2022mulan}, PER, and boundary MuLan similarity. We further introduce a dual-stage sampling strategy that fixes preceding segments with teacher-forced prefixes while resampling only the target segment and its subsequent context, enabling localized causal attribution for preference construction. This process yields approximately 140k song-, boundary-, and segment-level preference pairs for SAMP-DPO training.

% 为训练 SAMP-DPO，我们进一步构建了局部化的多粒度偏好数据集。针对每首歌曲随机采样生成八个候选版本，并基于 SongBench、段落级 MuLan 相似度、发音错误率（PER）以及边界 MuLan 相似度进行评估。随后，我们引入两阶段采样策略：利用 teacher-forced 前缀固定前序段落，仅对目标段落及其后续上下文进行重采样，从而实现偏好构建过程中的局部因果归因。最终获得约 14 万组整曲级、边界级与段落级偏好对，用于 SAMP-DPO 的训练。

\subsection{Experimental Setup}

\textbf{Song Generation Pretraining.}
We first re-implement and pre-train the LeVo autoregressive backbone from scratch on the 1.3M-song corpus. Starting from the reproduced LeVo checkpoint, SongSculpt is further trained for 60k steps with a learning rate of $1\times10^{-4}$ on 64 NVIDIA H20 GPUs, jointly optimizing Segment-wise Plan Grounding, Track-Factorized Realization, and Boundary-aware Transition Modeling .

% 我们首先从头复现了 LeVo 自回归骨干网络，并在包含 130 万首歌曲的语料库上完成预训练。在此基础上，SongSculpt 以复现得到的 LeVo 检查点为初始化，在 64 张 NVIDIA H20 GPU 上进一步训练 6 万步,学习率设为 1×10−4；整个训练联合优化分段级规划落地（Segment-wise Plan Grounding）、音轨因子化生成（Track-Factorized Realization）与边界感知过渡建模（Boundary-aware Transition Modeling）三个模块。

\textbf{Preference Alignment.}
The pretrained model is further optimized using SAMP-DPO on the constructed 140k localized preference pairs for 4k steps with a learning rate of $1\times10^{-6}$ on 64 NVIDIA H20 GPUs. Unless otherwise specified, all reported results are obtained using the SAMP-DPO-aligned model.

% 在预训练模型的基础上，我们进一步采用 SAMP-DPO，在构建的 14 万条局部偏好对上进行对齐优化：共训练 4k 步，学习率设为 1×10−6，训练在 64 张 NVIDIA H20 GPU 上完成。除非另有说明，本文报告的所有实验结果均基于经 SAMP-DPO 对齐后的模型获得。

\begin{table*}[t]
\centering
\small
\setlength{\tabcolsep}{4pt}
\renewcommand{\arraystretch}{1.12}
\setlength{\aboverulesep}{0pt}
\setlength{\belowrulesep}{0pt} 

\begin{tabular}{lcccccc|cccccccc}
\toprule

&
\multicolumn{5}{c}{\textbf{AudioBox}}
&
&
\multicolumn{8}{c}{\textbf{SongBench}}
\\

\cmidrule(lr){2-6}
\cmidrule(l){8-15}

Method
& CE & CU & PC & PQ & Mean & PER$\downarrow$
& Mel. & Arr. & Mus. & Voc. & Inst. & Mix. & Struct. & Mean
\\

\midrule

SFT
&
7.79
& 8.06
& 6.33
& \textbf{8.49}
& \underline{7.66}
& \textbf{8.6}
&
\textbf{6.20}
& \textbf{6.75}
& \textbf{5.28}
& \underline{6.28}
& \textbf{6.89}
& \textbf{6.36}
& \textbf{6.04}
& \textbf{6.26}
\\

w/o STP
&
\textbf{7.87}
& \underline{8.07}
& \textbf{6.57}
& \underline{8.48}
& \textbf{7.75}
& 16.2
&
6.02
& 6.46
& \textbf{5.28}
& \textbf{6.33}
& 6.70
& 6.24
& 5.83
& 6.12
\\

w/o SPG
&
7.71
& \textbf{8.08}
& 5.99
& 8.41
& 7.55
& 28.9
&
5.84
& 6.42
& 4.90
& 5.77
& 6.29
& 6.15
& 5.79
& 5.88
\\

w/o TFR
&
\underline{7.81}
& 8.04
& 6.31
& \underline{8.48}
& \underline{7.66}
& \underline{13.9}
&
6.08
& 6.59
& \underline{5.27}
& 6.25
& 6.77
& 6.29
& \underline{5.98}
& 6.18
\\

w/o BTM
&
\underline{7.81}
& 8.04
& \underline{6.36}
& 8.45
& \underline{7.66}
& 23.3
&
\underline{6.13}
& \underline{6.65}
& \textbf{5.28}
& 6.27
& \underline{6.81}
& \underline{6.32}
& \textbf{6.04}
& \underline{6.21}
\\

\bottomrule

\end{tabular}

\caption{
Ablation study of SongSculpt before SAMP-DPO alignment.
STP denotes Structured Temporal Planning,
SPG denotes Segment-wise Plan Grounding,
TFR denotes Track-Factorized Realization,
and BTM denotes Boundary-aware Transition Modeling.
Best and second-best results are highlighted in \textbf{bold} and \underline{underline}, respectively.
}

\label{tab:ablation}
\end{table*}

\subsection{Evaluation Metrics}

\textbf{Objective Metrics.}
Following recent song generation benchmarks, we employ Audiobox-Aesthetics~\cite{tjandra2025audiobox} to assess perceptual audio quality from four complementary perspectives: content enjoyment, content usefulness, production complexity, and production quality. We further adopt SongBench~\cite{songbench2025} to evaluate professional musical quality across multiple dimensions, including vocal quality, instrumental quality, melody, arrangement, musicality, mixing, and song structure. Finally, we report the Pronunciation Error Rate (PER) for lyric singing accuracy, segment-level MuLan similarity~\cite{huang2022mulan} for fine-grained instruction following, and boundary MuLan similarity for cross-segment continuity.

% 遵循近期歌曲生成基准的评测设置，我们采用 Audiobox-Aesthetics 从四个互补的视角评估音频的感知质量：内容愉悦度、内容有用性、制作复杂度与制作质量。在此基础上，我们进一步采用 SongBench 从多个维度评估专业音乐质量，涵盖人声质量、器乐质量、旋律、编曲、音乐性、混音与歌曲结构。最后，我们报告三项针对性指标：发音错误率（Pronunciation Error Rate, PER），用于衡量歌词演唱的准确性；分段级 MuLan 相似度，用于评估模型的细粒度指令遵循能力；边界 MuLan 相似度，用于衡量跨分段的连贯性。

\textbf{Subjective Metrics.}
Following the evaluation protocol adopted by recent song generation systems, we conduct human evaluation on six aspects using a five-point Likert scale: \emph{Overall Musical Quality}, \emph{Vocal \& Accompaniment Quality}, \emph{Vocal--Accompaniment Harmony}, \emph{Song Structure Clarity}, \emph{Lyric Accuracy}, and \emph{Instruction Faithfulness}. The first three metrics evaluate overall generation quality, while the latter three specifically assess structural planning, lyric rendering, and controllability, which are the primary objectives of SongSculpt. 

% 遵循近期歌曲生成系统所采用的评测协议，我们基于 five-point Likert scale，从以下六个方面开展人工评测：整体音乐质量（Overall Musical Quality）、人声与伴奏质量（Vocal & Accompaniment Quality）、人声–伴奏和谐度（Vocal–Accompaniment Harmony）、歌曲结构清晰度（Song Structure Clarity）、歌词准确性（Lyric Accuracy）以及指令遵循度（Instruction Faithfulness）。其中，前三项指标用于评估整体生成质量；后三项则分别针对结构规划、歌词呈现与可控性进行专项考察，而这三个方面正是 SongSculpt 的核心目标。

\subsection{Quantitative Results}

Table~\ref{tab:objective_results} shows that SongSculpt achieves state-of-the-art performance among all open-source models on both AudioBox and SongBench, while approaching the commercial system Suno V5.5. It also obtains the lowest PER, indicating substantially improved lyric rendering. Notably, Muse exhibits an unusually high PER because its generated songs often contain repetitive sections that lead to repeated lyric transcriptions by AceStep-1.5.

% 表 1 显示，SongSculpt 在 AudioBox 和 SongBench 两项评测上均取得了所有开源模型中的最优性能，并接近商用系统 Suno V5.5 的水平。同时，SongSculpt 获得了最低的 PER，表明其歌词呈现质量有显著提升。值得注意的是，Muse 的 PER 异常偏高，这是由于其生成的歌曲常包含重复段落，导致 AceStep-1.5 在转写时产生重复的歌词文本。

Compared with the SFT-only variant, SAMP-DPO consistently improves both AudioBox and SongBench while reducing PER from 8.6 to 4.5, demonstrating that localized preference optimization enhances overall musical quality without sacrificing controllability.

% 与仅使用 SFT 的变体相比，SAMP-DPO 在 AudioBox 和 SongBench 上均带来一致的性能提升，并将 PER 从 8.6 降至 4.5，表明局部化偏好优化能够在不牺牲可控性的前提下增强整体音乐质量。

As shown in Table~\ref{tab:segment_results}, SongSculpt further surpasses Muse in both average segment-level MuLan similarity and boundary continuity. The largest gains appear in vocal-intensive sections, while Intro, Instrumental, and Outro exhibit comparable scores. This trend is consistent with the proposed Track-Factorized Realization, which explicitly disentangles vocal and accompaniment generation, making fine-grained semantic control more effective in vocal-dominant segments. Moreover, SongSculpt achieves noticeably higher SongBench arrangement and structure scores, suggesting that Structured Temporal Planning further improves long-range musical organization beyond local controllability.

% 如表 2 所示，SongSculpt 在片段级 MuLan 平均相似度与边界连续性两项指标上均进一步超越 Muse。最大的提升出现在人声密集的段落，而前奏（Intro）、器乐段（Instrumental）与尾奏（Outro）的得分大体相当。这一趋势与本文提出的音轨分解式实现（Track-Factorized Realization）相一致：该设计将人声与伴奏的生成显式解耦，使细粒度语义控制在人声主导的片段中更为有效。此外，SongSculpt 在 SongBench 的编排与结构维度上取得了明显更高的分数，表明结构化时序规划在提升局部可控性之外，还进一步改善了长程的音乐组织。

\subsection{Human Evaluation}

Table~\ref{tab:subjective_results} reports the human evaluation results. SongSculpt achieves the best performance among all open-source models across all six dimensions and approaches the commercial system Suno V5.5, demonstrating that the proposed framework improves not only objective metrics but also overall perceptual quality. 

% 表 3 给出了人工评测结果。SongSculpt 在全部六个维度上均取得开源模型中的最佳表现，并接近商用系统 Suno V5.5 的水平，表明所提出的框架不仅提升了客观指标，也改善了整体感知质量。

The improvements are also consistent with the proposed design. Higher scores on \emph{Song Structure Clarity} validate the effectiveness of Structured Temporal Planning in organizing long-range musical evolution, while gains in \emph{Vocal \& Accompaniment Quality} and \emph{Vocal--Accompaniment Harmony} demonstrate the benefit of Track-Factorized Realization for coordinated vocal-accompaniment generation. More importantly, SongSculpt achieves substantially higher \emph{Instruction Faithfulness}, reflecting its stronger controllability through explicit plan-conditioned generation. Since the subjective experiment explicitly evaluates instruction following and structural controllability in addition to overall quality, it provides a more comprehensive assessment of controllable song generation than conventional music-quality evaluation alone.

% 这些改进与本文的设计思路高度吻合。在 \emph{Song Structure Clarity}（歌曲结构清晰度）上的更高得分，验证了结构化时序规划（Structured Temporal Planning）在组织长程音乐演进方面的有效性；而在 \emph{Vocal & Accompaniment Quality}（人声与伴奏质量）和 \emph{Vocal--Accompaniment Harmony}（人声—伴奏和谐度）上的提升，则体现了音轨分解式实现（Track-Factorized Realization）对于协调人声与伴奏生成的益处。更重要的是，SongSculpt 在 \emph{Instruction Faithfulness}（指令遵循度）上取得了显著更高的得分，反映出其通过显式的规划条件生成（plan-conditioned generation）所获得的更强可控性。由于该主观评测协议在整体质量之外还显式考察了指令遵循能力与结构可控性，因此与仅关注音乐质量的传统评测相比，它能够对可控歌曲生成提供更为全面的评估。

\subsection{Ablation Studies}

Table~\ref{tab:ablation} presents the ablation results before SAMP-DPO alignment. Removing Structured Temporal Planning (STP) mainly degrades SongBench, particularly arrangement and structure, indicating that explicit planning primarily improves long-range musical organization. In contrast, removing Segment-wise Plan Grounding (SPG) causes the largest overall degradation and increases PER from 8.6 to 28.9, demonstrating that faithfully grounding structured plans into autoregressive generation is essential for both controllability and lyric rendering.

% 表 4 展示了 SAMP-DPO 对齐之前的消融实验结果。移除结构化时序规划（STP）主要导致 SongBench 上的性能下降，尤其在编排与结构维度上最为明显，表明显式规划主要改善的是长程音乐组织。相比之下，移除逐片段计划落地（SPG）造成了最大幅度的整体性能退化，并使 PER 从 8.6 升至 28.9，说明将结构化计划忠实地落地到自回归生成过程中，对可控性与歌词呈现均至关重要。

Track-Factorized Realization (TFR) and Boundary-aware Transition Modeling (BTM) provide complementary improvements. TFR consistently enhances melody, instrumentation, and overall musicality through track-disentangled generation, while BTM improves structural consistency and cross-segment transitions. Overall, the results validate that explicit planning, faithful plan grounding, and structured realization are all indispensable to SongSculpt.

% 分轨解耦生成（TFR）与边界感知过渡建模（BTM）则带来了互补性的提升：TFR 通过分轨解耦的生成方式持续改善旋律、器乐编配与整体音乐性，而 BTM 增强了结构一致性与跨片段过渡的质量。总体而言，上述结果验证了显式规划、忠实的计划落地以及结构化的实现对 SongSculpt 均不可或缺。

\section{Conclusion}

In this paper, we present SongSculpt, a plan-conditioned framework for fine-grained controllable song generation. By introducing an explicit Structured Temporal Plan that disentangles global intentions from segment-level evolution, SongSculpt enables interpretable and editable control. Experiments demonstrate state-of-the-art performance among open-source song generation systems with significantly improved fine-grained controllability, highlighting a promising paradigm for controllable song generation.

%本文提出了 SongSculpt，一种面向细粒度可控歌曲生成的规划条件生成框架。通过在自回归生成之前引入显式的 Structured Temporal Plan，将全局意图与段落级演进进行解耦，SongSculpt 实现了可解释、可编辑的生成控制。大量实验表明，SongSculpt 在开源歌曲生成系统中取得了当前最优性能，并显著提升了细粒度可控能力，验证了可控歌曲生成新范式的有效性。

\bibliography{aaai2027}


\end{document}